\documentclass[journal]{IEEEtran}
\usepackage{amsmath,amssymb,amsfonts}
\usepackage{graphicx}
\usepackage{textcomp}
\usepackage{subfig}
\usepackage{xcolor}
\usepackage{caption}
\usepackage{cite}
\usepackage{listings,xcolor}
\usepackage[utf8]{inputenc}
\usepackage{float}
\usepackage{adjustbox}
\usepackage{etoolbox}
\usepackage{pifont}

\usepackage{framed}
\usepackage{url}
\usepackage{pbox}
\usepackage{flushend}

\usepackage[hidelinks]{hyperref}
\hypersetup{hidelinks}

\usepackage{diagbox,color}
\usepackage{multirow}
\usepackage{tabularx}
\usepackage{makecell}

\usepackage{hyperref}

\begin{document}

\title{ An Evolutionary Pathway for the Quantum Internet Relying on Secure Classical Repeaters}

\author{Gui-Lu Long, Dong Pan, Yu-Bo Sheng, Qikun Xue, Jianhua Lu, and Lajos Hanzo
\thanks{Gui-Lu Long is with the State Key Laboratory of Low-Dimensional Quantum Physics and Department of Physics, Tsinghua University, 
and the Frontier Science Center for Quantum Information, Beijng 100084, China, 
and the Beijing Academy of Quantum Information Sciences, Beijing 100193, China.}

\thanks{Dong Pan is with the Beijing Academy of Quantum Information Sciences, Beijing 100193, China 
and the State Key Laboratory of Low-Dimensional Quantum Physics and Department of Physics, Tsinghua University, Beijing 100084, China.}

\thanks{Yu-Bo Sheng is with the College of Electronic and Optical Engineering, Nanjing University of Posts and Telecommunications, Nanjing, 210003, China. }

\thanks{Qikun Xue is with the State Key Laboratory of Low-Dimensional Quantum Physics and Department of Physics, Tsinghua University, 
the Frontier Science Center for Quantum Information, Beijng 100084, China, 
the Beijing Academy of Quantum Information Sciences, Beijing 100193, China, 
and the Southern University of Science and Technology, Shenzhen 518055, China.}

\thanks{Jianhua Lu is with the School of Information Science and Technology, Tsinghua University, 
the Beijing National Research Center for Information Science and Technology, 
and the Frontier Science Center for Quantum Information, Beijing 100084, China.}

\thanks{Lajos Hanzo is is with the School of Electronics and Computer Science, University of Southampton, Southampton SO17 1BJ, United Kingdom.}
}
\maketitle

\begin{abstract}
Until quantum repeaters become mature, quantum networks remain restricted either to limited areas of directly connected nodes or to nodes connected to a common node. We circumvent this limitation by conceiving quantum networks using secure classical repeaters combined with the quantum secure direct communication (QSDC) principle, which is a compelling form of quantum communication that directly transmits information over quantum channel. The final component of this promising solution is our classical quantum-resistant algorithm. Explicitly, in these networks, the ciphertext gleaned from a quantum-resistant algorithm is transmitted using QSDC along the nodes, where it is read out and then transmitted to the next node. At the repeaters, the information is protected by our quantum-resistant algorithm, which is secure even in the face of a quantum computer. Hence, our solution offers secure end-to-end communication across the entire network, since it is capable of both eavesdropping detection and prevention in the emerging quantum internet. It is compatible with operational networks, and will enjoy the compelling services of the popular Internet, including authentication. Hence, it smoothens the transition from the classical Internet to the Quantum Internet (Qinternet) by following a gradual evolutionary upgrade. It will act as an alternative network in quantum computing networks in the future. We have presented the first experimental demonstration of a secure classical repeater based hybrid quantum network constructed by a serial concatenation of an optical fiber and free-space communication link. In conclusion, secure repeater networks may indeed be constructed using existing technology and continue to support a seamless evolutionary pathway to the future Qinternet of quantum computers. 
\end{abstract}

\section{Introduction}
\label{s1-introduction}
The Quantum internet (Qinternet) links nodes such as quantum computers, quantum sensors as well as other quantum devices, and it will support fascinating functions and tasks that are impossible in the classical Internet \cite{ref1-lloyd2004infrastructure}. A quantum network includes nodes, routers, repeaters, and quantum channels \cite{ref3-wehner2018quantum}. A quantum link includes fiber or free-space optical channels, and the specific type of communications protocol adopted, such as quantum secure direct communication (QSDC) \cite{ref5-long2002theoretically}, quantum key distribution (QKD) \cite{ref4-bennett1984quantum}, or quantum teleportation \cite{ref6-bennett1993teleporting}. 

The roadmap to build the fully-fledged Qinternet has been envisioned in \cite{ref3-wehner2018quantum}. According to its functionality, six stages lie ahead. At the time of writing, the first three stages are in the experimental phase, and the trusted-repeater networks concept has the potential of extending the network to large areas. It forms an important stepping stone towards the Qinternet \cite{ref3-wehner2018quantum}, although it does not provide end-to-end security. Prepare-and-measure networks provide links for directly-linked nodes, while entanglement-distribution networks link up nodes connected to a common intermediate node, where the qubits received from the other two nodes are measured using the measurement-device-independent technique of \cite{ref10-zhou2020measurement}. Thus, the scale of quantum networks classified as prepare-and-measure networks and entanglement-distribution networks remain limited. Satellite-based repeaters are capable of increasing the distance between the directly connected nodes, but they remain limited to 3 nodes, so they are also at the entanglement-distribution network stage.

Once having practical quantum memories becomes a reality, quantum repeaters will be used in quantum networks for extending their scale. However, it is still a tremendous open challenge to fabricate practical quantum memory. Furthermore, the incompatibility of quantum communication with the existing classical Internet presents another challenge. Therefore, the Qinternet remains unable to exploit the vast resources and powerful facilities of the conventional Internet.

Against this background, we propose the secure-repeater networks (SRNs) concept by intrinsically combining QSDC \cite{ref5-long2002theoretically,ref7-deng2003two,ref8-deng2004secure} and classical cryptographic algorithms, in particular, the family of quantum-resistant algorithms~\cite{ref9-alagic2020status}. In these networks, the ciphertext gleaned from a quantum-resistant algorithm is transmitted using QSDC step-by-step along the nodes, where it is read out and retransmitted to the next node.

In contrast to relying solely on the trust in the nodes of trusted-repeater networks, computational security \footnote{A cryptosystem can be said to be computationally secure when the computational requirements become so excessive that it becomes practically infeasible to break the cryptosystem in 'reasonable time'.} is provided at the classical repeater nodes. Thus it provides secure end-to-end communication for the entire network. Communication satisfying information-theoretic security \footnote{A cryptosystem can be said to be information-theoretically secure when its security can be purely proven by information theory, and the cryptosystem cannot be broken even if the adversary has unlimited computing power. } is provided using QSDC for directly connected nodes, or when using measurement-device-independent QSDC \cite{ref10-zhou2020measurement} for the nodes connected to a common node, respectively. Because the nodes of SRNs are linked by QSDC, both eavesdropping detection and prevention capabilities are inherent in SRNs.

In contrast to the conventional Internet, SRN replaces classical communication by quantum communication. Since SRNs are fully compatible with the existing Internet, they circumvent the incompatibility challenge in the development of the Qinternet. As a benefit, SRNs can take full advantage of the compelling facilities and services provided by the operational Internet, above all, its salient authentication function. The compatibility of SRNs creates an evolutionary path for the Qinternet. However, it is worth noting that SRNs are not merely a provisional measure conceived for circumventing the lack of quantum memory. Quite the opposite, the services of SRNs are still required even at further evolved stages of the Qinternet, making it a potent alternative network for the future Qinternet.

This paper is organized as follows. In Sec.~\ref{s2-qsdc}, we introduce the basic concept of QSDC, which is an essential ingredient of SRNs. In Sec.~\ref{s3-srn}, we describe the SRN stage, while in Sec.~\ref{s4-7stages}, we present a roadmap of constructing the Qinternet in 7 evolutionary stages. The corresponding proof of concept demonstration is provided in Sec.~\ref{s5-demonstration}. Finally, a summary is given in Sec. \ref{s6-summary}.

\section{Quantum secure direct communication}
\label{s2-qsdc}
\begin{figure}[!h]
\begin{center}
\includegraphics[width=8.8cm,angle=0]{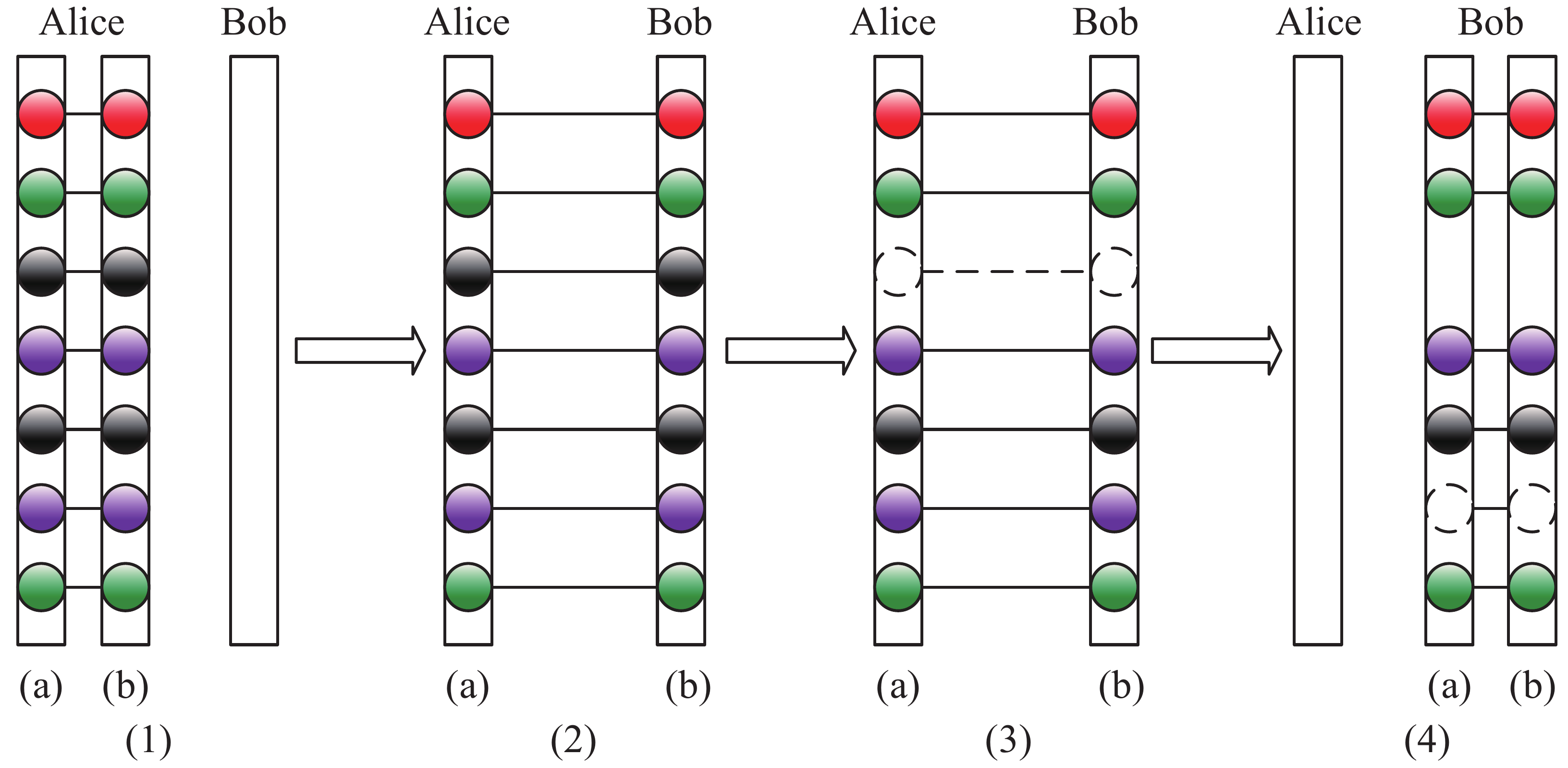}
\caption{Procedure of the efficient-QSDC protocol \cite{ref5-long2002theoretically}.}
\label{fig1-efficientqsdc}
\end{center}
\end{figure}

QSDC was originally proposed in 2000 \cite{ref5-long2002theoretically}, which transmits information directly through a quantum channel, while maintaining information-theoretic security without a prior secret key distribution stage. QSDC can be realized using entanglement \cite{ref5-long2002theoretically,ref7-deng2003two} or using single-photons \cite{ref8-deng2004secure}. Here we briefly highlight the first QSDC protocol \cite{ref5-long2002theoretically}, where information is mapped to the quantum states of the Einstein-Podolsky-Rosen (EPR) pairs. The illustration of the protocol is shown in Fig. \ref{fig1-efficientqsdc}, which contains the following steps:

(1) Alice prepares an ordered sequence of EPR-pairs containing both the message bits and check bits. Explicitly, first she prepares a sequence of EPR-pairs according to the message. Then she also chooses some random numbers mapped to the checking bits and prepares the corresponding EPR-pairs, which are inserted into the message EPR-pair sequence at random positions. As shown in Fig. \ref{fig1-efficientqsdc}, the pair of spheres labeled by red, green, black and purple represent the entangled state that carries information or checking bits 00, 01, 10 and 11, respectively. Then Alice chooses from each EPR-pair one qubit to form the ordered sequence (b), and the remaining qubits form another ordered single qubit sequence (a).

(2) Alice sends the (b)-sequence to Bob.

(3) Alice measures some of the checking qubits using either the Z or the X bases randomly. She then informs Bob of the positions, of the basis and the results of these measurements through a classical channel. Bob then measures the corresponding particles in the (b)-sequence at his site. He then continues by comparing his results with Alice's results and estimates the error rate. If it is below a threshold, then they proceed to the next step. Otherwise the process is abandoned and restarted. The pair of dotted circles in Fig. \ref{fig1-efficientqsdc} represents the checking qubits.

(4) In case of successful confirming the security of the transmission of the (b)-sequence, Alice sends the (a)-sequence to Bob. He applies Bell-basis measurements to the received EPR-pairs and reads out both the message bits and the remaining checking bits. Alice proceeds by announcing the states of the remaining checking bits, followed by Bob comparing them to his results. If the error rate is below the tolerable threshold, they conclude that the transmission is successful.

This scheme is deemed to be an efficient-QSDC protocol, since it harnesses all EPR pairs for transmitting information - except for those selected for checking the presence of eavesdroppers \cite{ref5-long2002theoretically}.

In the protocol of Fig. \ref{fig1-efficientqsdc}, information is mapped to the quantum states of the EPR-pairs, while in \cite{ref7-deng2003two}, information is mapped to the dense coding operations applied to the EPR-pairs. A single-photon QSDC protocol has also been proposed in \cite{ref8-deng2004secure}. However, the original QSDC protocols still rely on quantum memory to wait for the results of eavesdropping checking. As a remedy, recently, quantum-memory-free QSDC protocols were proposed, where the information is protected by classical coding, which serendipitously plays the role of quantum memory \cite{ref12-9130765}. The security analysis of QSDC has been completed using Wyner's wiretap channel theory \cite{ref14-wyner1975wire,ref13-qi2019implementation}. As an evolution from classical communications, where infinitesimally low error probability is guaranteed by Shannon's theory in noisy channels, QSDC \cite{ref5-long2002theoretically,ref7-deng2003two,ref8-deng2004secure} simultaneously guarantees both reliable and secure communication in the presence of both noise and eavesdropping. Recently, a practical QSDC prototype has been constructed \cite{ref13-qi2019implementation}, and its upgraded version is capable of communicating at 4 kbps over a 10 km fiber link, which is eminently suitable for texts, images, and telephone calls.

Table \ref{table:QKDvsQSDC} provides a comparison of QSDC and QKD from the perspectives of their history, functionality, security, communication pattern, and their best performance at the time of writing. QKD was proposed earlier than QSDC and it has been fully developed. It can be seen that the practical performance of QSDC is below that of QKD at the time of writing. Hence, the engineering of QSDC must be accelerated. The SRN network concept proposed here can be relied upon for mitigating the security risks imposed by trusted repeater, since it is capable of supporting end-to-end secure communication at computational security relying on quantum-resistant algorithms.

\begin{table*}
\centering
\caption{Comparison between QKD and QSDC.}
\label{table:QKDvsQSDC}
\begin{tabular}{|l|m{7cm}|m{7cm}|} 
\hline
\multicolumn{1}{|c|}{\diagbox{Features}{Schemes}} & \multicolumn{1}{c|}{QKD}                                                                                                                                                                                               & \multicolumn{1}{c|}{QSDC}                                                                                                                                                                                                                                                                                          \\ 
\hline
History                                          & The first QKD protocol was proposed by C. H. Bennett and G. Brassard in 1984.                                                                                                      & The first QSDC protocol was proposed by G. L. Long and X. S. Liu in 2000.                                                                                                                                                                                                      \\ 
\hline
Functionality          
                          
& \makecell[{{m{7cm}}}]{Negotiating secret key between a pair of legitimate users.
 \\ $\bullet$ \textit{Pros}: Easy to implement with privacy amplification for generation of the final secret key. 
 \
 \\ $\bullet$ \textit{Cons}: Limited application; Need key management. \\~}
& \makecell[{{m{7cm}}}]{Realizing reliable and secure communication directly between a pair of legitimate users. 
\\$\bullet$ \textit{Pros}: Wide application, secure communication of information and key without explicit encryption.\
 \\$\bullet$ \textit{Cons}: More complicated to implement compared to QKD.}   \\
\hline
Security                                         
& \makecell[{{m{7cm}}}]{Security is assured by detecting eavesdropping and discarding transmitted data are leaked. \\$\bullet$ \textit{Pros}: Information-theoretically secure; Supporting measurement-device-independent protocol and device-independent protocol; Eavesdropping detection.\\~ \\$\bullet$ \textit{Cons}: Lack of industrial standards for engineering; Security risks in trusted repeater.}                                                                                                   
&  \makecell[{{m{7cm}}}]{Security is assured by detecting and preventing eavesdropping. \\$\bullet$ \textit{Pros}: Information-theoretically secure; Supporting measurement-device-independent protocol and device-independent protocol; Eavesdropping detection; Eavesdropping prevention; Secure repeater offers secure end-to-end communication. \\$\bullet$ \textit{Cons}: Lack of industrial standards for engineering.\\~}                                                                                                                                                                                 \\ 
\hline
Pattern                                          & QKD negotiates secure key first, information is transmitted through another classical communication.                                                                                                                         & QSDC completes reliable and secure communication in a single transmission.                                                                                                                                                      \\ 
\hline
Best performance                                   
& \makecell[{{m{7cm}}}]{
$\bullet$ Secret key rate: $>$2 Mbps @10 km fiber link.\\~\\~}                                                                                                                                       & \makecell[{{m{7cm}}}]{
$\bullet$ Information rate: 4.85 kbps @ 12.04 km fiber link. \\$\bullet$Theoretically, it has the same distance and rate as QKD using current technology.}                                                                                                                                                                                                                            \\
\hline
\end{tabular}
\end{table*}

\section{secure repeater aided quantum networks}
\label{s3-srn}
The SRN is a quantum network relying on QSDC transmitting ciphertext generated by a classical cryptographic algorithm, such as a quantum-resistant algorithm \cite{ref9-alagic2020status}. At a node of the SRN, the ciphertext is read out and then routed to the next node using QSDC again. The information is protected by classical cryptography benefiting from computational security both at the node as well as for end-to-end communication across the whole network. For directly-linked nodes, communication capable of maintaining information-theoretic security is provided using QSDC. For a pair of nodes connected to a common node, measurement-device-independent QSDC \cite{ref10-zhou2020measurement} can be used, which also provides information-theoretic security. Naturally, in the case of two nodes connected to a common node, a secure classical repeater can also be used for maintaining computational security, which is easier to implement than measurement-device-independent QSDC.

The security principles of quantum and classical cryptography are distinctly different. Instead of viewing them as competing technologies, they may be viewed as complementary to each other in SRNs. QSDC provides security by denying the eavesdropper's chance to intercept information, either plain text or ciphertext. Hence an eavesdropper fails to decrypt any useful information, even if he/she has abundant computing power. By contrast, classical cryptography relies on solving excessively complex mathematical problems - such as the factorization of large numbers - for ensuring that Eve cannot solve the problem in a reasonably short time. The combination of these two technologies in SRNs is capable of further increasing the security level. Explicitly, this limits the potential acquisition of ciphertext from the entire Internet to some repeater nodes. Furthermore, the eavesdropping detection capability of legitimate users will spot any illegal eavesdropping activities, and will enhance protection in those areas, where eavesdropping is frequently detected.

\begin{figure}[!h]
\begin{center}
\includegraphics[width=8.7cm,angle=0]{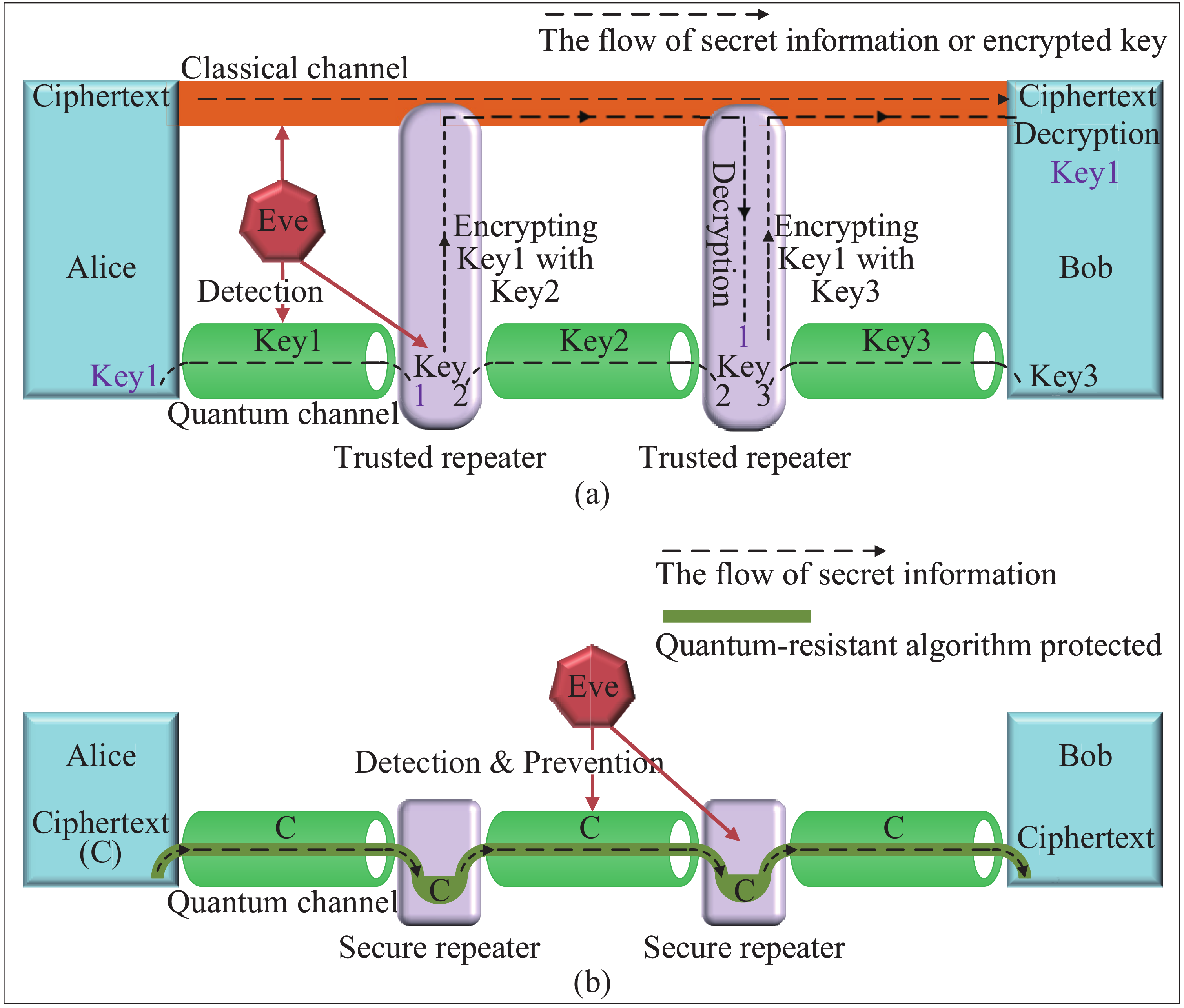}
\caption{The model of (a) trusted repeater network and (b) secure repeater network.}
\label{fig2-TRN-SRN}
\end{center}
\end{figure}

SRN provides secure end-to-end communication, in contrast to trusted-repeater network, where the latter provides no end-to-end security.  Although both trusted-repeater network and SRN use classical repeaters, there is an essential difference between them. As shown in Fig. \ref{fig2-TRN-SRN} (a), in trusted-repeater network, the key 'key1' is negotiated between Alice and node 1, key 'key2'  between node 1 and node 2, and so on until the destination is reached.  Key 'key1' is used between the transmitter and the receiver. In trusted-repeater network, 'key1' will be relayed from node 1 encrypted using a classical cryptographic algorithm, say the one-time-pad using key 'key2', to node 2, then to node 3 encrypted with 'key3' until the receiver of Bob is reached. Thus, 'key1' is known to all the intermediate repeater nodes, and they must be trusted for ensuring that the key 'key1' will not leak. Compared to existing classical networks, where the information is protected by classical cryptography in both the communication channels and repeater nodes, the security of transmission in trusted-repeater network is enhanced. However, the security level at the intermediate nodes is reduced to relying solely on the trust of these nodes. In contrast to trusted-repeater networks, in SRNs, the information is encrypted using a classical cryptographic algorithm, and it is protected by relying on computational security at the repeater nodes, as shown in Fig. \ref{fig2-TRN-SRN} (b). It should be pointed out that the same strategy cannot be used in trusted-repeater networks, because the random nature of the negotiated key prohibits the pre-encryption of the key.

The substitution of classical communication across the classical Internet by employing QSDC in SRNs has the benefit of both eavesdropping detection and prevention capabilities, which are unique to QSDC. In SRNs, at each channel transmission of QSDC between a pair of nodes, we may incorporate a random bit sequence as the label of each transmitted packet, and publicize them after the packet reaches the end user. By comparing the positions of bit errors, the legitimate users can readily spot the specific channel, where eavesdropping takes place. By exploiting these capabilities, both the users and service providers can deploy enforced protection in geographic areas, where eavesdropping is frequently detected. This beneficial capability will increase the security level of SRNs. This is particularly important, because the unconditional security of classical cryptographic algorithms, including quantum-resistant algorithms, has not been unequivocally proven. By contrast, the unconditional security of quantum communication has been proven in principle. When quantum computing networks have finally been realized, information-theoretic security will be realized across the entire Qinternet and SRNs will pave the way for this process. Both trusted-repeater network and SRN require an authenticated classical channel for exchanging data concerning eavesdropping detection in QKD and QSDC, respectively, which were omitted in Fig. \ref{fig2-TRN-SRN}. The classical channel required for quantum-resistant algorithm also has been omitted in Fig. \ref{fig2-TRN-SRN} (b), because it constitutes a pre-processing stage for SRN-based communication.

\section{Seven-stage roadmap of the Qinternet evolution}
\label{s4-7stages}

\begin{figure}[!h]
\begin{center}
\includegraphics[width=8.7cm,angle=0]{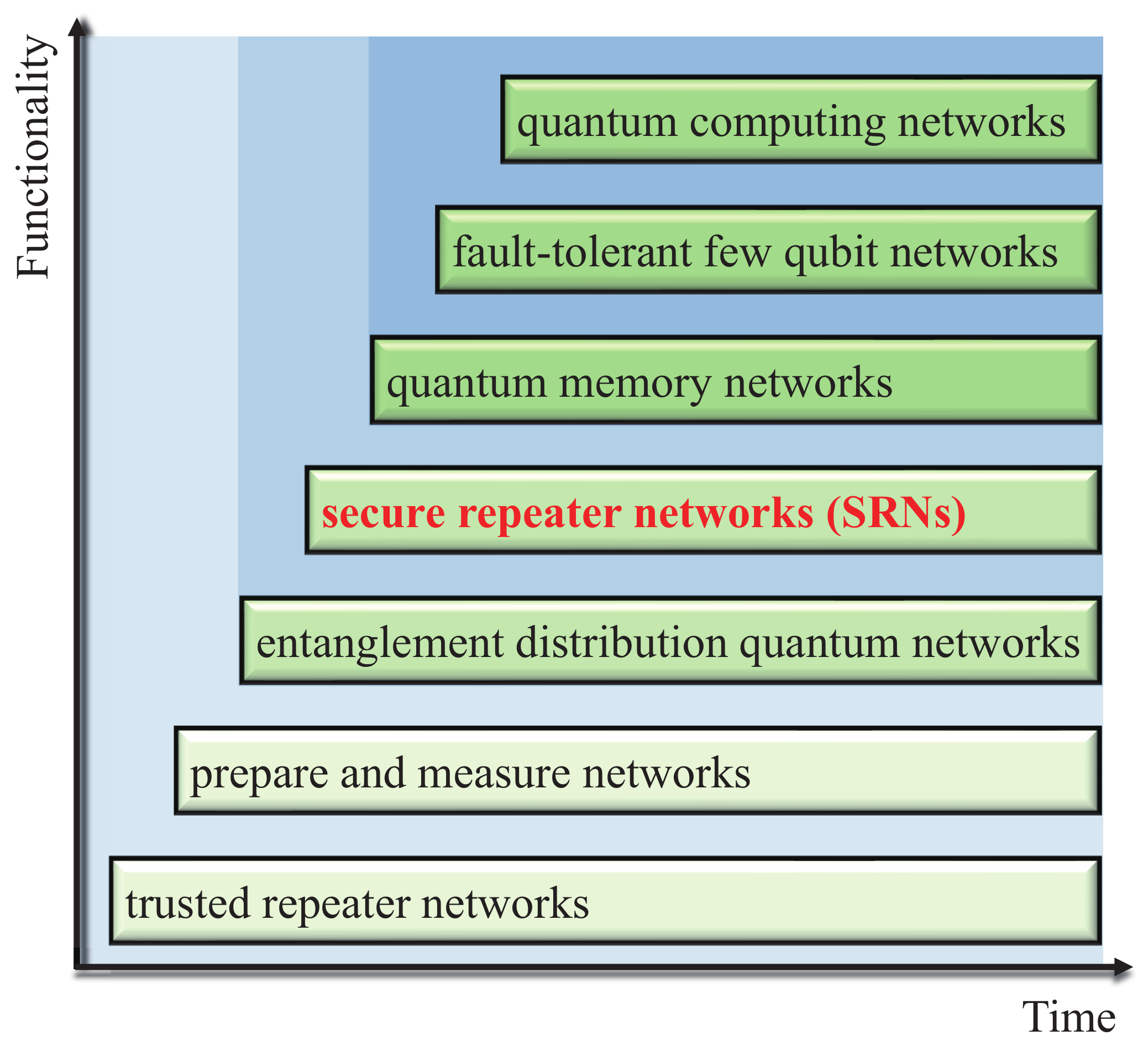}
\caption{The seven-stage Qinternet evolved from the six-stage scheme of S. Wehner {\em et al.}~\cite{ref3-wehner2018quantum} by adding secure repeater networks.}
\label{fig3-7stages}
\end{center}
\end{figure}

Let us now evolve the Qinternet roadmap outlined by S. Wehner \textit{et al.} \cite{ref3-wehner2018quantum} a little further by incorporating the SRN stage, as seen in Fig. \ref{fig3-7stages}. The definitions and the specific types of tasks in each of the other six stages were detailed in~\cite{ref3-wehner2018quantum}. 

Each stage has its own application examples, which are listed below.
\begin{itemize}
\item[] \textbf{Stage 1 - trusted repeater networks}: QKD with no end-to-end security can be constructed relying on the trusted-repeater networks.
\item[] \textbf{Stage 2 - Prepare-and-measure networks}: The prepare-and-measure networks support QKD, QSDC and secure identification.
\item[] \textbf{Stage 3 - entanglement-distribution networks}: The entanglement-distribution networks constitute the foundations of device independent protocols.
\item[] \textbf{Stage 4 - secure repeater networks}:  The SRNs are compatible with the operational Internet and provide secure end-to-end communication relying on computational security. However, compared to the Internet purely relying on quantum-resistant algorithms, an SRN adds both eavesdropping detection as well as prevention for the evolved Internet. Furthermore, it can serve as an auxiliary network of the ultimate fully-fledged Qinternet.
\item[] \textbf{Stage 5 - quantum memory networks}: Blinding quantum computing becomes available by relying on quantum memory networks.
\item[] \textbf{Stage 6 - fault-tolerant few qubit networks}: Both clock synchronization and distributed quantum computation are the products of the fault-tolerant few qubit networks.
\item[] \textbf{Stage 7 - quantum computing networks}: The quantum computing networks afford leader election and prompt Byzantine agreement.
\end{itemize}

Long-distance end-to-end communication could also be achieved using hop-by-hop QSDC in prepare-and-measure networks relying on trusted intermediate nodes in the same way as in trusted repeater networks. This evolutionary step has been facilitated by the proposed SRN concept for the first time. SRNs fill the gap between entanglement-distribution networks, and quantum memory networks, which might need decades to develop, hence SRNs accelerate the evolution of quantum networks. In the context of the seven-stage roadmap, SRN  has a pair of important roles.

To elaborate, SRN only requires replacing classical communication by quantum communication. The immediate advantage for SRN users is that they can enjoy all the services of the operational Internet, such as authentication and packet switching based transmission. Asymmetric cryptography such as quantum-resistant algorithm provides efficient authentication in the operational Internet at a complexity which increases linearly with the number of users. By contrast, in quantum cryptography, any pair of users has to pre-share a common secret in order to perform authentication. Hence the associated complexity becomes intractable, because it increases quadratically with the number of users.

Given the huge benefits of interconnecting quantum computers only capable of processing a modest number of qubits for constructing more powerful quantum computers, the Qinternet must be rolled out in the near future. However, the time-scale of rolling out the Qinternet is uncertain. Because SRNs are compatible with the classical Internet, we can start from SRN at the earliest and then may flexibly update parts of the SRN upon replacing the secure classical repeaters by quantum repeaters. While updating the system, the SRN and the accompanying Qinternet operating either in fault-tolerant few qubit networks or in quantum computing networks will continue to operate simultaneously, making a gradual and smooth transition to the full Qinternet. It is expected that this kind of hybrid quantum networks will exist for several decades.

The SRN will continue to exist as an alternative network even at the quantum computing network stage, because it is capable of supporting the functionalities of the classical Internet, such as authentication, certification and so on. The DL04 QSDC-based SRN is also capable of exchanging an arbitrary known single qubit. Both quantum repeaters and classical secure repeaters co-exist at the nodes. SRNs are also capable of finding the route from the source to the destination, while relying on QSDC.

\section{Proof of concept demonstration of SRN}
\label{s5-demonstration}
\begin{figure}[!h]
\begin{center}
\includegraphics[width=8.9cm,angle=0]{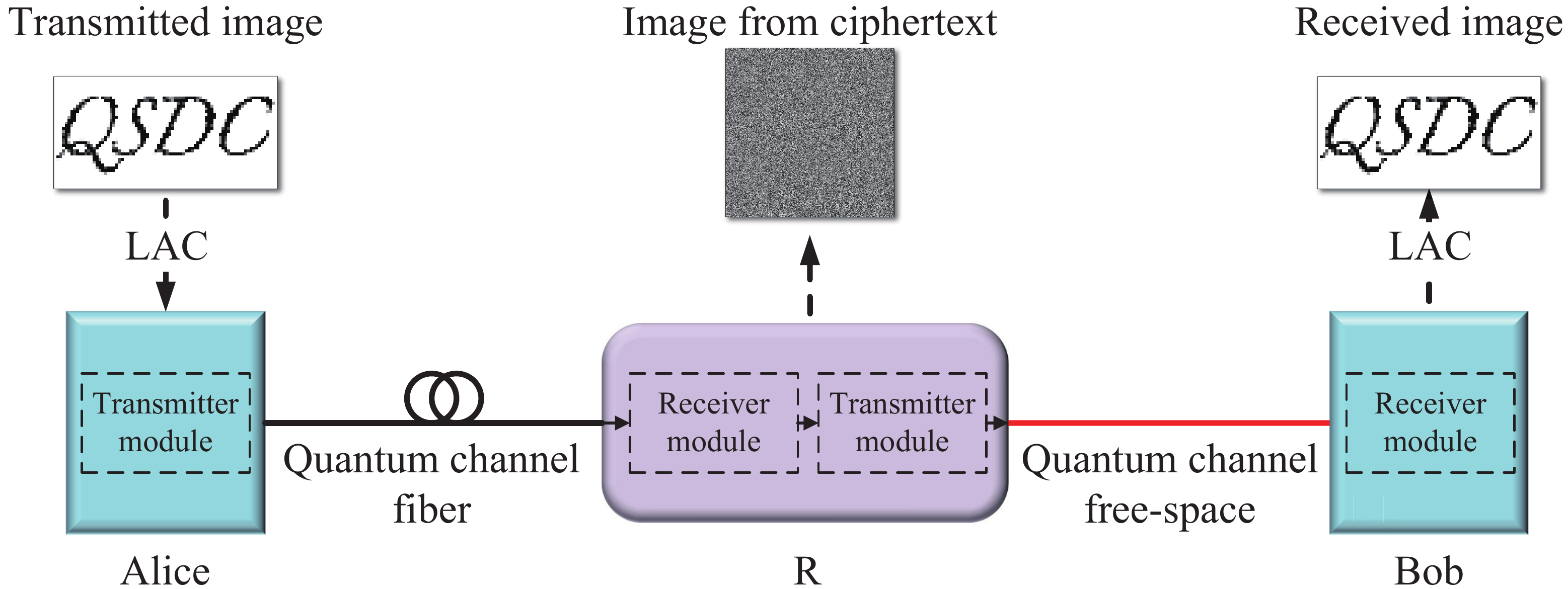}
\caption{The SRN demonstration system. LAC is the abbreviation of a LAttice based Cryptosystem.}
\label{fig4-ExpDem}
\end{center}
\end{figure}

As shown in Fig.~\ref{fig4-ExpDem}, a three-node SRN is constructed,
where nodes Alice and R are connected through a 10 km fiber link, while
nodes R and Bob are connected by a 2-meter free-space link~\cite{pan2020experimental}. Node R plays
the role of a secure-classical repeater. We have adopted a lattice based cryptosystem, LAC, a
quantum resistant algorithm designed by X. H. Lu \textit{et al.}, a
candidate quantum-resistant algorithm in the 2nd round post quantum cryptography competition of NIST
\cite{ref9-alagic2020status}. Lattice-based cryptographic enjoys very strong security based on worst-case hardness. In LAC, a 32-byte information string is encrypted into 1024 bytes of ciphertext. In this demonstration an image file is transformed into an 800-kbyte encrypted file and transmitted from Alice to the Relay R and then to Bob. The DL04 QSDC~\cite{ref8-deng2004secure} is adopted for both hops transmissions. Explicitly, the single photons transmitted from Alice are detected at the receiver module of R for recovering the classical bits of the ciphertext, which are then mapped again to new single photons sent to Bob by using the QSDC transmitter module. Upon reception it
is then decrypted to recover the original image. At the repeater node,
the encrypted image is undecipherable by the repeaters, as it
transpires from Fig.~\ref{fig4-ExpDem}. These experiments demonstrate
the reliable operation of the SRN.

\begin{figure}[!h]
\begin{center}
\includegraphics[width=8.7cm,angle=0]{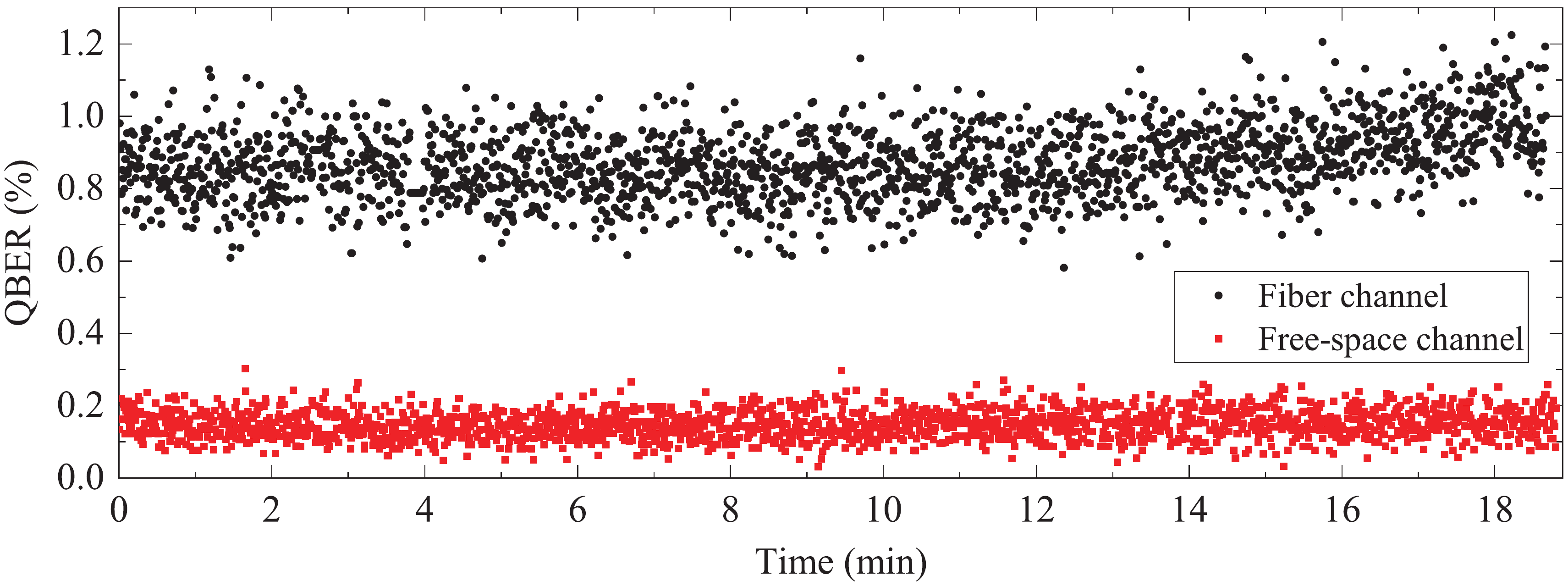}
\caption{The channel quantum bit error rates during image file transmission.}
\label{fig5-QBER}
\end{center}
\end{figure}

The pair of QSDC links are fully functional, relying on low density parity check coding
and secure transmission. Due to concatenating a pair of QSDC links,
it takes slightly longer to transmit the encrypted file from R to Bob,
than from Alice to R. The channel's quantum bit error rates are
also different and both of them are below the error rate threshold 5.7\%, as shown in Fig.~\ref{fig5-QBER}. The higher quantum bit error rate of fiber channel than that of the free space result arises from its higher length-related attenuation. There are a whole
host of interesting networking issues set aside for further study,
such as the associated throughputs and delays, just to name a few.

\section{Conclusions and outlooks}
\label{s6-summary}
In a nutshell, we have proposed an evolutionary pathway for the Qinternet relying on secure classical repeaters, the SRN. It relies on QSDC that transmits ciphertext encrypted by classical quantum resistant algorithms. In contrast to purely relying on the trust of the repeater nodes in trusted-repeater networks, it guarantees having computational security. It provides secure end-to-end communication for the entire quantum network right away at the current state-of-the-art. It has eavesdropping detection and prevention capabilities for the Internet, which mitigates the risk of future retroactive decryption currently faced by classical cryptographic algorithms, when higher computing power becomes available. It is compatible with the existing Internet and hence it can exploit all the functions provided by the Internet,  which paves the way for a smooth and gradual transition to the Qinternet. It will serve as an ancillary network for the ultimate Qinternet, providing authentication, routing and other Internet services. In conclusion, we have constructed a three-node SRN testbed and demonstrated the secure transmission of an image encrypted by the quantum-resistant algorithm over our concatenated fiber and free-space quantum channel.

Finally, it is worth mentioning that QSDC - which is the central core of SRN - is gradually developing into a perfectly secure practical application. The SRN will secure next-generation communications and will support the creation of long-haul quantum networks, once QSDC becomes fully developed \cite{you2021towards}. A working QSDC prototype that transmits 4 kbps of secure information through a 10 km fiber link was reported in \cite{ref13-qi2019implementation}. The distance and transmission rate can be further increased by using the technology of increasing capacity using masking \cite{long2021drastic}, and it is estimated that a 300 kbps information rate can be achieved over a 50 km fiber link in the near future, and the technology is also suitable for satellite applications, making SRN concept an implementation-technology at the time of writing. 

\section*{Acknowledgement}
The authors would like to thank Dr. Zengrong Zhou as well as Dr. Shijie Wei for their help in the quantum-resistant algorithm LAC and acknowledge the helpful discussions with Prof. Liuguo Yin. This work was supported in part by the National Natural Science Foundation of China (Grant No. 11974205 and No. 11974189), in part by the National Key Research and Development Program of China under Grant No. 2017YFA0303700 and in part by the Key R\&D Program of Guangdong Province (Grant No. 2018B030325002). L. Hanzo would like to acknowledge the financial support of the Engineering and Physical Sciences Research Council projects EP/P034284/1 and EP/P003990/1 (COALESCE) as well as of the European Research Council's Advanced Fellow Grant QuantCom (Grant No. 789028).

\bibliographystyle{IEEEtran}
\bibliography{ref}


\vskip -2\baselineskip plus -1fil

\begin{IEEEbiographynophoto}{Gui-Lu Long} [M'16] (gllong@mail.tsinghua.edu.cn), FAPS, FIOP, received his B.S. degree in 1982 and his Ph.D. degree in 1987. He is a professor at Tsinghua University. He served as President of AAPPS (2017-2019), and Vice-chair of C13 of IUPAP (2015-2017). He published 300+ refereed papers, and has 21000+ citations.\end{IEEEbiographynophoto}

\vskip -2\baselineskip plus -1fil

\begin{IEEEbiographynophoto}{Dong Pan} (pand16@tsinghua.org.cn) received his B.S. degree from the Northwest University in 2016 and his Ph.D. degree from the Tsinghua University in 2021. From 2018 to 2019, he was a Visiting Student with the University of Southampton. He is an assistant research scientist at Beijing Academy of Quantum Information Sciences.
\end{IEEEbiographynophoto}

\vskip -2\baselineskip plus -1fil

\begin{IEEEbiographynophoto}{Yu-bo Sheng} (shengyb@njupt.edu.cn) received his Ph.D. degree from Beijing Normal University in 2009 and engaged in postdoctoral research in Tsinghua University from 2009 to 2011. He is a professor at Nanjing University of Posts and Telecommunications. His research interests are quantum communication, quantum repeater, and quantum computation.
\end{IEEEbiographynophoto}

\vskip -2\baselineskip plus -1fil

\begin{IEEEbiographynophoto}{Qikun Xue} (qkxue@tsinghua.edu.cn) FAPS, received his B.S. degree in 1984 and his Ph.D. degree in 1994. He is a member of Chinese Academy of Sciences. He won the Fritz London Memorial Prize for low-temperature physics in 2020. He has authored/coauthored \textasciitilde370 papers. 
\end{IEEEbiographynophoto}

\vskip -2\baselineskip plus -1fil
\begin{IEEEbiographynophoto}{Jianhua Lu} [F'15] (lhh-dee@mail.tsinghua.edu.cn), received his B.S. and M.S. degrees from Tsinghua University in 1986 and 1989, respectively, and his Ph.D. degree from the Hong Kong University of Science \& Technology in 1998. He is now a vice president of NSFC, a member of Chinese Academy of Sciences.
\end{IEEEbiographynophoto}

\vskip -2\baselineskip plus -1fil

\begin{IEEEbiographynophoto}{Lajos Hanzo} [F'04] (lh@ecs.soton.ac.uk) (http://www-mobile.ecs.soton.ac.uk, https://en.wikipedia.org/wiki/Lajos\_Hanzo), received Honorary Doctorates  from the Technical University of Budapest and Edinburgh University. He is a Foreign Member of the Hungarian Science-Academy, Fellow of the Royal Academy of Engineering (FREng), of the IET, of EURASIP and holds the Eric Sumner Field Award.
\end{IEEEbiographynophoto}

\end{document}